\documentstyle[12pt,epsfig]{article}

\renewcommand{\thefootnote}{\fnsymbol{footnote}}

\newcounter{sectionc}\newcounter{subsectionc}\newcounter{subsubsectionc}
\renewcommand{\section}[1] {\vspace*{0.6cm}\addtocounter{sectionc}{1}
\setcounter{subsectionc}{0}\setcounter{subsubsectionc}{0}\noindent
	{\normalsize\bf\thesectionc. #1}\par\vspace*{0.4cm}}
\renewcommand{\subsection}[1] {\vspace*{0.6cm}\addtocounter{subsectionc}{1}
	\setcounter{subsubsectionc}{0}\noindent
	{\normalsize\it\thesectionc.\thesubsectionc. #1}\par\vspace*{0.4cm}}
\renewcommand{\subsubsection}[1]
{\vspace*{0.6cm}\addtocounter{subsubsectionc}{1}
	\noindent {\normalsize\rm\thesectionc.\thesubsectionc.\thesubsubsectionc.
	#1}\par\vspace*{0.4cm}}

\newcounter{appendixc}
\newcounter{subappendixc}[appendixc]
\newcounter{subsubappendixc}[subappendixc]

\renewcommand{\appendix}[1] {\vspace*{0.6cm}
        \refstepcounter{appendixc}
        \setcounter{figure}{0}
        \setcounter{table}{0}
        \setcounter{equation}{0}
        \renewcommand{\thefigure}{\Alph{appendixc}.\arabic{figure}}
        \renewcommand{\thetable}{\Alph{appendixc}.\arabic{table}}
        \renewcommand{\theappendixc}{\Alph{appendixc}}
        \renewcommand{\theequation}{\Alph{appendixc}.\arabic{equation}}
        \noindent{\bf Appendix \theappendixc #1}\par\vspace*{0.4cm}}

\def\abstracts#1{{

\centering{\begin{minipage}{12.2truecm}\footnotesize\baselineskip=12pt\noindent
	\centerline{\footnotesize ABSTRACT}\vspace*{0.3cm}
	\parindent=0pt #1
	\end{minipage}}\par}}

\renewenvironment{thebibliography}[1]
	{\begin{list}{\arabic{enumi}.}
	{\usecounter{enumi}\setlength{\parsep}{0pt}
\setlength{\leftmargin 1.25cm}{\rightmargin 0pt}
	 \setlength{\itemsep}{0pt} \settowidth
	{\labelwidth}{#1.}\sloppy}}{\end{list}}

\newcounter{itemlistc}
\newcounter{romanlistc}
\newcounter{alphlistc}
\newcounter{arabiclistc}

\newcommand{\tcaption}[1]{
        \refstepcounter{table}
        \setbox\@tempboxa = \hbox{\footnotesize Table~\thetable. #1}
        \ifdim \wd\@tempboxa > 6in
           {\begin{center}
        \parbox{6in}{\footnotesize\baselineskip=12pt Table~\thetable. #1}
            \end{center}}
        \else
             {\begin{center}
             {\footnotesize Table~\thetable. #1}
              \end{center}}
        \fi}

 1
 1
 1

\font\ninerm=cmr9

\catcode`\@=11
\long\def\@makefntext#1{
\protect\noindent \hbox to 3.2pt {\hskip-.9pt
$^{{\ninerm\@thefnmark}}$\hfil}#1\hfill}		

\def\@makefnmark{\hbox to 0pt{$^{\@thefnmark}$\hss}}  

\def\ps@myheadings{\let\@mkboth\@gobbletwo
\def\@oddhead{\hbox{}
\rightmark\hfil\ninerm\thepage}
\def\@oddfoot{}\def\@evenhead{\ninerm\thepage\hfil
\leftmark\hbox{}}\def\@evenfoot{}
\def\sectionmark##1{}\def\subsectionmark##1{}}

\def\nostrocostrutto#1\over#2{\mathrel{\mathop{\kern 0pt \rlap
  {\raise.2ex\hbox{$#1$}}}
  \lower.9ex\hbox{\kern-.190em $#2$}}}

\newcommand{\be}{\begin{equation}}
\newcommand{\ee}{\end{equation}}
\newcommand{\ba}{\begin{eqnarray}}
\newcommand{\ea}{\end{eqnarray}}

\newcommand{\eref}[1]{(\ref{#1})}      
\newcommand{\N}{{\mathcal N}}

\begin{document}

\newcommand{\fcaption}[1]{
        \refstepcounter{figure}
        \setbox\@tempboxa = \hbox{\footnotesize Fig.~\thefigure. #1}
        \ifdim \wd\@tempboxa > 6in
           {\begin{center}
        \parbox{6in}{\footnotesize\baselineskip=12pt Fig.~\thefigure. #1}
            \end{center}}
        \else
             {\begin{center}
             {\footnotesize Fig.~\thefigure. #1}
              \end{center}}
        \fi}

\def\@citex[#1]#2{\if@filesw\immediate\write\@auxout
	{\string\citation{#2}}\fi
\def\@citea{}\@cite{\@for\@citeb:=#2\do
	{\@citea\def\@citea{,}\@ifundefined
	{b@\@citeb}{{\bf ?}\@warning
	{Citation `\@citeb' on page \thepage \space undefined}}
	{\csname b@\@citeb\endcsname}}}{#1}}

\newif\if@cghi
\def\cite{\@cghitrue\@ifnextchar [{\@tempswatrue
	\@citex}{\@tempswafalse\@citex[]}}
\def\citelow{\@cghifalse\@ifnextchar [{\@tempswatrue
	\@citex}{\@tempswafalse\@citex[]}}
\def\@cite#1#2{{$\null^{#1}$\if@tempswa\typeout
	{IJCGA warning: optional citation argument
	ignored: `#2'} \fi}}
\newcommand{\citeup}{\cite}

\thispagestyle{empty}
\rightline{MPI-PhT/95-94}
\rightline{October 1995}

\vspace{2.5cm}

\centerline{\Large \bf Moment analysis of energy spectra}
\vspace{0.2cm}
\centerline{\Large\bf and the effect of running coupling\footnote{to be
published in the Proceedings of the XXV$^{th}$ International Symposium
on Multiparticle Dynamics, Star\'a
Lesn\'a, Slovakia, September 12-16,  1995, Eds. D. Brunsko, L. Sandor, J.
Urban,
World Scientific, Singapore}}

\vspace{1.cm}

\centerline{\large Sergio Lupia\footnote{E-mail address: lupia@mppmu.mpg.de}}

\vspace{1.cm}

\centerline{\it Max-Planck-Institut
f\"ur Physik, Werner-Heisenberg-Institut}
\vspace{0.1cm}
\centerline{\it F\"ohringer Ring 6, D--80805 M\"unchen, Germany}

\vspace{1.0cm}

\begin{abstract}
Single particle inclusive energy spectra in $e^+e^-$ annihilation
are analyzed in terms of moments.
By assuming Local Parton Hadron Duality (LPHD),
experimental data in a wide c.m. energy range from 3 GeV up to LEP energy
are  compared to the theoretical predictions of
Modified Leading Log Approximation (MLLA) of QCD with and without taking into
account the running of $\alpha_s$.
MLLA with running coupling  (Limiting Spectrum)
is found to reproduce experimental results very well, while the model with
fixed coupling is inconsistent with data.
Rescaled cumulants are shown to be sensitive to the running of $\alpha_s$
in the asymptotic regime, while the
Lorentz-invariant distribution, $E dn/d^3p$, points out this effect at
very small energy $E$ of few hundreds MeV.
These results give a direct evidence of the running  of the QCD coupling
in inclusive energy spectra and lend further support to the LPHD picture.
\end{abstract}

\newpage

\setcounter{page}{1}

\centerline{\normalsize\bf MOMENT ANALYSIS OF ENERGY SPECTRA}
\baselineskip=22pt
\centerline{\normalsize\bf AND THE EFFECT OF RUNNING COUPLING}

\vspace*{0.6cm}
\centerline{\footnotesize SERGIO LUPIA}
\baselineskip=13pt
\centerline{\footnotesize\it Max-Planck-Institut f\"ur
Physik, Werner-Heisenberg-Institut}
\baselineskip=12pt
\centerline{\footnotesize\it F\"ohringer Ring 6, D-80805 M\"unchen, Germany}
\centerline{\footnotesize E-mail: lupia@mppmu.mpg.de}
\vspace*{0.3cm}
\vspace*{0.9cm}
\abstracts{Single particle inclusive energy spectra in $e^+e^-$ annihilation
are analyzed in terms of moments.
By assuming Local Parton Hadron Duality (LPHD),
experimental data in a wide c.m. energy range from 3 GeV up to LEP energy
are  compared to the theoretical predictions of
Modified Leading Log Approximation (MLLA) of QCD with and without taking into
account the running of $\alpha_s$.
MLLA with running coupling  (Limiting Spectrum)
is found to reproduce experimental results very well, while the model with
fixed coupling is inconsistent with data.
Rescaled cumulants are shown to be sensitive to the running of $\alpha_s$
in the asymptotic regime, while the
Lorentz-invariant distribution, $E dn/d^3p$, points out this effect at
very small energy $E$ of few hundreds MeV.
These results give a direct evidence of the running  of the QCD coupling
in inclusive energy spectra and lend further support to the LPHD picture.}

\normalsize\baselineskip=15pt
\setcounter{footnote}{0}
\renewcommand{\thefootnote}{\alph{footnote}}

\section{Introduction}

The main open problem in strong interaction phenomenology is the understanding
of the hadronization mechanism. There is indeed a gap between theoretical
predictions coming from perturbative QCD at parton level and experimental
results at final particle level.
A  bridge between these two sides is given by a soft
confinement mechanism\cite{BAS,DKMTbook}:
single particle inclusive spectra are simply required to be proportional to
single parton inclusive spectra obtained from perturbative QCD, provided
the parton cascade is evolved down to a
cutoff scale of the order of the hadronic masses
$Q_0\simeq m_h$.
This hypothesis of Local Parton Hadron Duality (LPHD)\cite{LPHD} has been
particularly successful in describing both the shape and the position of the
maximum of single inclusive charged particle momentum spectra in hadronic jets
in $e^+e^-$ annihilation ranging from PETRA/PEP\cite{tasso} to
LEP energy\cite{opal,l3,aleph,delphi}.
Recently, similar results have been obtained in DIS for inclusive momentum
spectra in the Breit frame at HERA\cite{zeus,H1}.

In this paper, I report on the results obtained in collaboration with Wolfgang
Ochs\cite{brussels}. We performed the analysis of moments up to order 4 of
single
particle inclusive energy spectra in hadronic jets in $e^+e^-$ annihilation.
Since moments are sensitive to the tail of the distribution, i.e., to the low
energy region, a consistent kinematical scheme has been defined to
relate parton and hadron spectra.
Our aim was two-fold: firstly, to investigate the sensitivity of single
particle  energy spectra  to the running  of the
coupling $\alpha_s$, in particular for soft particles in the low energy region,
where the coupling is stronger; secondly, to test the validity of the
description in terms of perturbative QCD plus LPHD in the small c.m. energy
domain.

\section{The theoretical framework}

\subsection{MLLA Evolution Equations}

The evolution equation of single parton inclusive energy
distribution in the framework of MLLA is given by\cite{DKMTbook}:
\be
\frac{d}{d\log \Theta} x \bar D_A^B(x,\log P\Theta ) = \sum_{C=q,\bar q,g}
\int_0^1 dz \frac{\alpha_s(k_t)}{2 \pi} \Phi^C_A(z) \left[ \frac{x}{z} \bar
D^B_C \left( \frac{x}{z}, \log (z P \Theta) \right) \right]
\label{uno}
\ee
with the boundary condition at threshold $P \Theta = Q_0$,
$x \bar D_A^B(x,\log Q_0) = \delta(1-x) \delta_A^B$.
Here $P$ and $\Theta$ denote the primary parton
energy and jet opening angle respectively,
$x$ is the energy fraction carried by the produced parton,
$\Phi_i^j(z)$ are the  parton splitting kernels and $i$,$j$, $A$, $B$,
$C$ label quarks, antiquarks and gluons.
The QCD running coupling is given by its one-loop expression
$\alpha_s(k_t) = 2 \pi/ b \log (k_t / \Lambda)$ with
$b \equiv (11 N_c - 2 n_f)/3$,
$\Lambda$ the QCD-scale and  $N_c$ and $n_f$ the number of colors and of
flavors respectively.
The scale of the coupling is given by the transverse momentum
$k_t \simeq z (1-z) P \Theta$. The shower evolution is cut off by
$Q_0$, such that $k_t \ge Q_0$.

The integral equation can be solved by Mellin transform:
\be
D_{\omega}(Y,\lambda) = \int_0^1 \frac{dx}{x} x^{\omega}
[x \bar D(x,Y,\lambda)] =
\int_0^Y d\xi e^{-\xi \omega} D(\xi,Y, \lambda)
\ee
with
$Y = \log \frac{P \Theta}{Q_0} \simeq \log \frac{P}{Q_0}$,
$\lambda = \log \frac{Q_0}{\Lambda}$, $\xi = \log \frac{P}{k}$
and parton energy $k$.

In flavor space the valence quark and $(\pm)$ mixtures of sea quarks and gluons
evolve independently with different ``eigenfrequencies''. At high energies,
the dominant contribution to the inclusive spectrum comes from
the ``plus''-term,  which we denote by
$D_{\omega}(Y,\lambda) \equiv D^+_{\omega}(Y,\lambda)$.
By keeping now, as usual, only the leading singularity in $\omega$-space
plus a constant term, one gets:
\be
\label{master}
\left( \omega + \frac{d}{dY} \right) \frac{d}{dY} D_{\omega}(Y,\lambda)
 - 4 N_c \frac{\alpha_s(Y+\lambda)}{2 \pi} D_{\omega}(Y,\lambda)
= -  a \left( \omega + \frac{d}{dY} \right) \frac{\alpha_s(Y+\lambda)}{2 \pi}
D_{\omega}(Y,\lambda)
\ee
where  $a = 11 N_c / 3 + 2 n_f / 3 N_c^2$ accounts for recoil effects.

By defining the anomalous dimension $\gamma_{\omega}$ according to:
\be
D_{\omega}(Y,\lambda) = D_{\omega}(Y_0,\lambda) \exp \left( \int_{Y_0}^Y dy
\gamma_{\omega}[\alpha_s(y+\lambda)] \right)
\label{anom}
\ee
the evolution equation for the inclusive spectrum can also
be written as  a differential equation for $\gamma_{\omega}$.
In general this nonlinear equation has two roots and the solution consists of a
superposition of two terms of the type~\eref{anom}.
The explicit solution of eq.~\eref{master} can be expressed indeed
as a linear superposition of two hypergeometric distributions\cite{DKMTbook}.
An interesting case of the general solution
is the Limiting Spectrum, where the two parameters coincide,
$Q_0 = \Lambda$, i.e., $\lambda = 0$. In this case,
the expressions simplify  and a simpler integral representation of the
spectrum has been achieved.

\subsection{MLLA with fixed $\alpha_s$}

In order to study the sensitivity of  the single particle
inclusive energy distribution
to the running of the coupling, we built a new model based on
MLLA formalism but  with the coupling kept fixed at a given value.
In this case, the value of the coupling or, equivalently, the
value of the parameter $\gamma_0$,  $\gamma_0^2 = 6 \alpha_s / \pi$,
replaces $\Lambda$ as  a free parameter of the model.
Accordingly, eq.~\eref{master} becomes now:
\be
\left( \omega + \frac{d}{dY} \right) \frac{d}{dY} D_{\omega}(Y,\lambda) -
\gamma_0^2  D_{\omega}(Y,\lambda)
= -  2 \eta \left( \omega + \frac{d}{dY} \right) D_{\omega}(Y,\lambda)
\ee
with $\eta = a \gamma_0^2 / 8 N_c = a \alpha_s /4 \pi$.

One can then perform explicitly the inverse Mellin transform
and get an explicit solution for the inclusive energy spectrum in MLLA with
fixed-$\alpha_s$:
\be
D_{fix}(\xi,Y,\lambda) = \gamma_0 \sqrt{\frac{Y-\xi}{\xi}}
I_1 \biggl(2 \gamma_0 \sqrt{\xi (Y-\xi)} \biggr)  e^{- 2 \eta (Y-\xi)}
\label{fixed:solution}
\ee

\section{Moment analysis: from momentum spectra to energy spectra}

\subsection{Moment analysis}

In addition to the standard study of single particle inclusive distribution,
one can also study the multiplicity $\bar \N$ and the normalized
moments of the distribution, $<\xi^q(Y, \lambda)> $, given by:
\ba
\bar \N  &=& \int d\xi D(\xi,Y,\lambda)  \\
<\xi^q(Y, \lambda)> &=& \frac{1}{\bar \N}
\int d\xi \xi^q D(\xi,Y,\lambda) \nonumber
\ea
or the cumulant
moments $\kappa_q(Y,\lambda)$\cite{FW,DKTInt};
the latter are related to ordinary moments by  a standard cluster expansion:
\ba
\kappa_1 &=& <\xi> = \bar \xi \quad \quad , \quad
 \kappa_2 \equiv \sigma^2 = <(\xi - \bar \xi)^2> \\
 \kappa_3 &=& <(\xi - \bar \xi)^3> \quad , \quad
 \kappa_4 = <(\xi - \bar \xi)^4> - 3 \sigma^4 \nonumber
 \ea
One also introduces the reduced cumulants $k_q \equiv \kappa_q/ \sigma^q$, in
particular the skewness $s = k_3$ and the kurtosis $k = k_4$.
The cumulants $\kappa_q$ can be derived from the expansion:
\be
\log D_{\omega}(Y,\lambda) =
\sum_{q=0}^{\infty} \kappa_q(Y,\lambda) \frac{(- \omega)^q}{q!} \;
\Leftrightarrow \;
\kappa_q(Y,\lambda) =  \left( - \frac{\partial}{\partial \omega}
\right)^q \log D_{\omega}(Y,\lambda) \biggl|_{\omega=0} \; ,
\label{cumulants}
\ee
with $\bar \N_E(Y,\lambda) = D_{\omega}(Y,\lambda) \bigl|_{\omega = 0}$.

At high energies, one term of the form \eref{anom} dominates and one obtains
\be
\kappa_q(Y) = \kappa_q(Y_0) +
\int_{Y_0}^Y dy \left( - \frac{\partial}{\partial \omega}
\right)^q \gamma_{\omega}[\alpha_s(y)] \biggl|_{\omega=0}
\label{mom2}
\ee
This equation shows the direct dependence of the moments on $\alpha_s(Y)$. For
fixed $\alpha_s$, for example, one obtains directly $\kappa_q(Y) \propto Y$ for
high energies.

Let us stress that
the study of moments and cumulants is worth to be pursued, since explicit
analytical expressions for their dependence on c.m. energy have been
obtained\cite{DKTInt};
moments of order $q$  in the MLLA model with running coupling can indeed
be expressed as a function of
 $a$, $b$, $Y+\lambda$ and $\lambda$, while moments for the Limiting Spectrum
can be expressed in a simpler way  in terms of the parameter $B \equiv a/b$
and the variable $z \equiv \sqrt{ 16 N_c Y / b }$.
In the fixed $\alpha_s$ model, the solution is straightforward, since in this
case the evolution equation for the anomalous dimension reduces to an algebraic
equation.
In addition,
if the higher order cumulants ($q > 2$) are sufficiently small, one can
reconstruct the $\xi$-distribution from the first four order cumulants,
using a Distorted Gaussian parametrization of the spectrum\cite{FW}.

However, let us notice again
that high order moments are sensitive to the tail of the
distribution; therefore, one has to look carefully at the treatment of the low
energy region $E \to Q_0$. This point is discussed in detail in the next
Subsection.

\subsection{The rescaling procedure}

In theoretical calculations, partons are taken as
massless, $E_p = p_p$,  but  a $p_t$ cutoff $Q_0$, $p_t \ge Q_0$,
is needed for infrared  regularization;  therefore, $E_p \ge Q_0$ and the
theoretical spectrum  goes linearly to 0 for $E \to Q_0$.
These predictions are usually compared to experimental results on single
particle inclusive momentum spectrum, which show  a finite tail down to
very low momentum (large $\xi$).
The difference of the two spectra in their tails
 could affect significantly the moment analysis.
In order to cure this anomaly, let us remind that observable hadrons are
massive, therefore $E_h \ge m_h$. It has been  therefore suggested to identify
the
cutoff parameter $Q_0$ appearing in theoretical calculations with an effective
hadron mass averaged over all charged particles.
In addition, one has to look
at energy spectra rather than at momentum spectra.
Let us also remind that the Lorentz-invariant distribution
$E\frac{dn}{d^3p}$  approximately follows an exponential law in low-energy
experiments for sufficiently small energies:
\be
E\frac{dn}{d^3p} \sim e^{-E/E_0}
\label{exp}
\end{equation}
with $E_0 \simeq$ 150 MeV\cite{dasp,adone,slac}.
By performing a suitable rescaling, the spectrum $E_h dn/dp_h$ goes then
linearly to 0 for $E \to Q_0$.

In conclusion, we propose to compare theoretical predictions for single
particle
inclusive energy spectra with the experimental spectrum
\be
E_h \frac{dn(\xi_E)}{dp_h} \ \ \hbox{vs.} \ \ \xi_E  = \log \frac{1}{x_E} =
\log
\frac{P}{E}
\label{dual}
\ee
where an effective  mass $Q_0$ is assigned to all particles, i.e.,
$E_h = \sqrt{p_h^2+Q^2_0}$. Notice  then that experimental spectra do
depend on the chosen value of $Q_0$.

\section{Results}

\subsection{Moments}

Figure~1 shows (the logarithm of)
the average multiplicity $\bar {\mathcal N}_E$ and the first four
cumulants of charged particle energy spectra
as a function of $Y = \log (\sqrt{s}/2Q_0)$ for $Q_0$ = 270 MeV.
For the calculation of moments, an additional value for the bin in the
unmeasured interval near $\xi_E \simeq Y$ (small momenta) has been added
by linear  extrapolation.
Errors of the moments are taken as the sum of statistical and systematic
errors.
The latter particularly affects the overall normalization and then the average
multiplicity; higher order moments are on the contrary independent of the
overall normalization.
The errors of the moments also includes the errors on the central values
of $\xi_E$ in each bin, taken as half the bin-size.
%
%
The value of $Q_0$ has been obtained by comparing
the moments determined for a selected $Q_0$ with the theoretical predictions
of MLLA with running $\alpha_s$  at different values of $\Lambda$.
The best agreement is obtained for the Limiting Spectrum
\be
Q_0 = \Lambda \simeq 270 \pm 20 \  \hbox{MeV}
\ee
Predictions of the Limiting Spectrum  with this  value of $Q_0$
and number of flavors $n_f$ = 3 are shown in the Figure (solid lines).
\begin{figure}
          \begin{center}
          \mbox{
\mbox{\epsfig{file=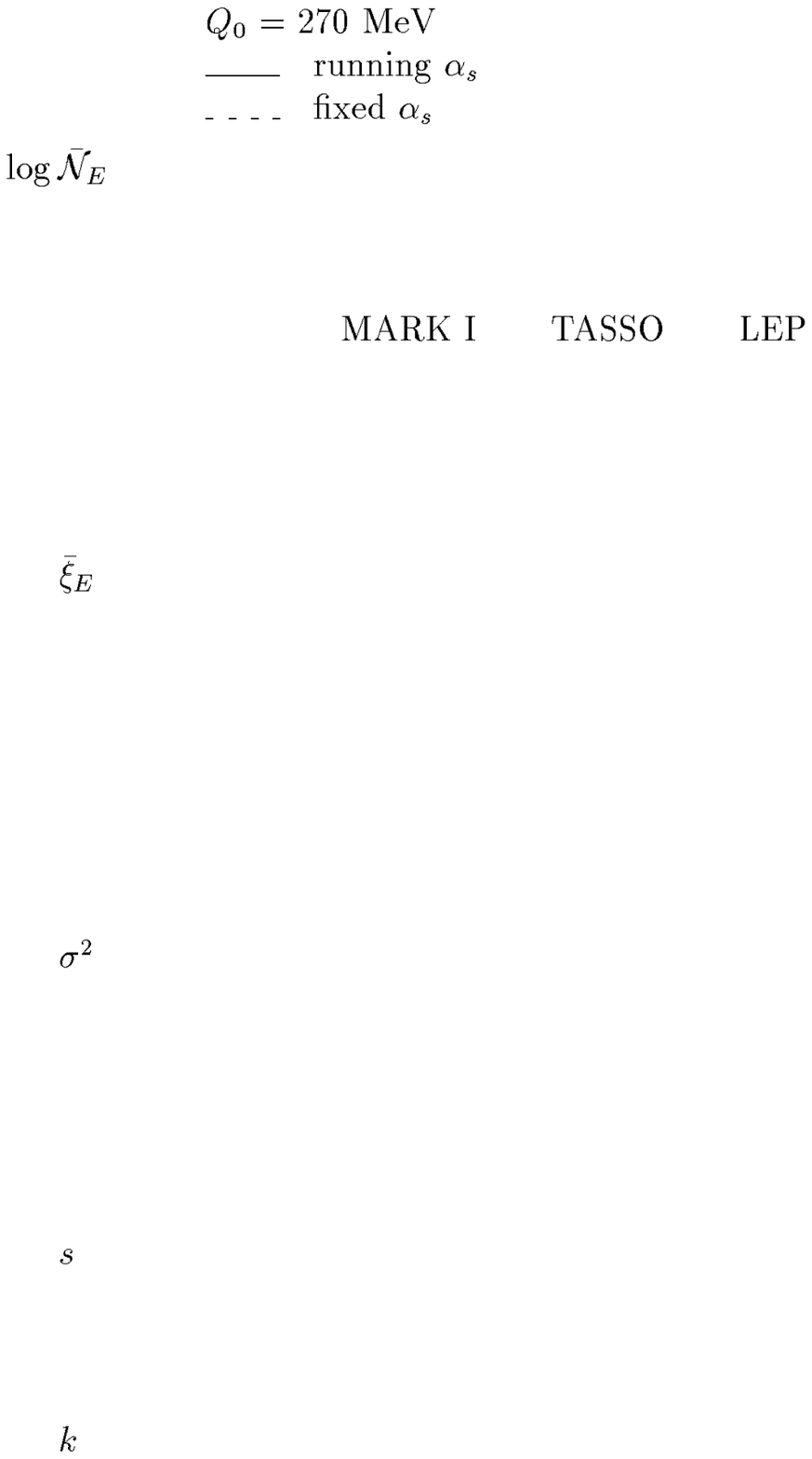,bbllx=5.cm,bblly=7.cm,bburx=5.2cm,bbury=26.cm,height=16cm}}
\mbox{\epsfig{file=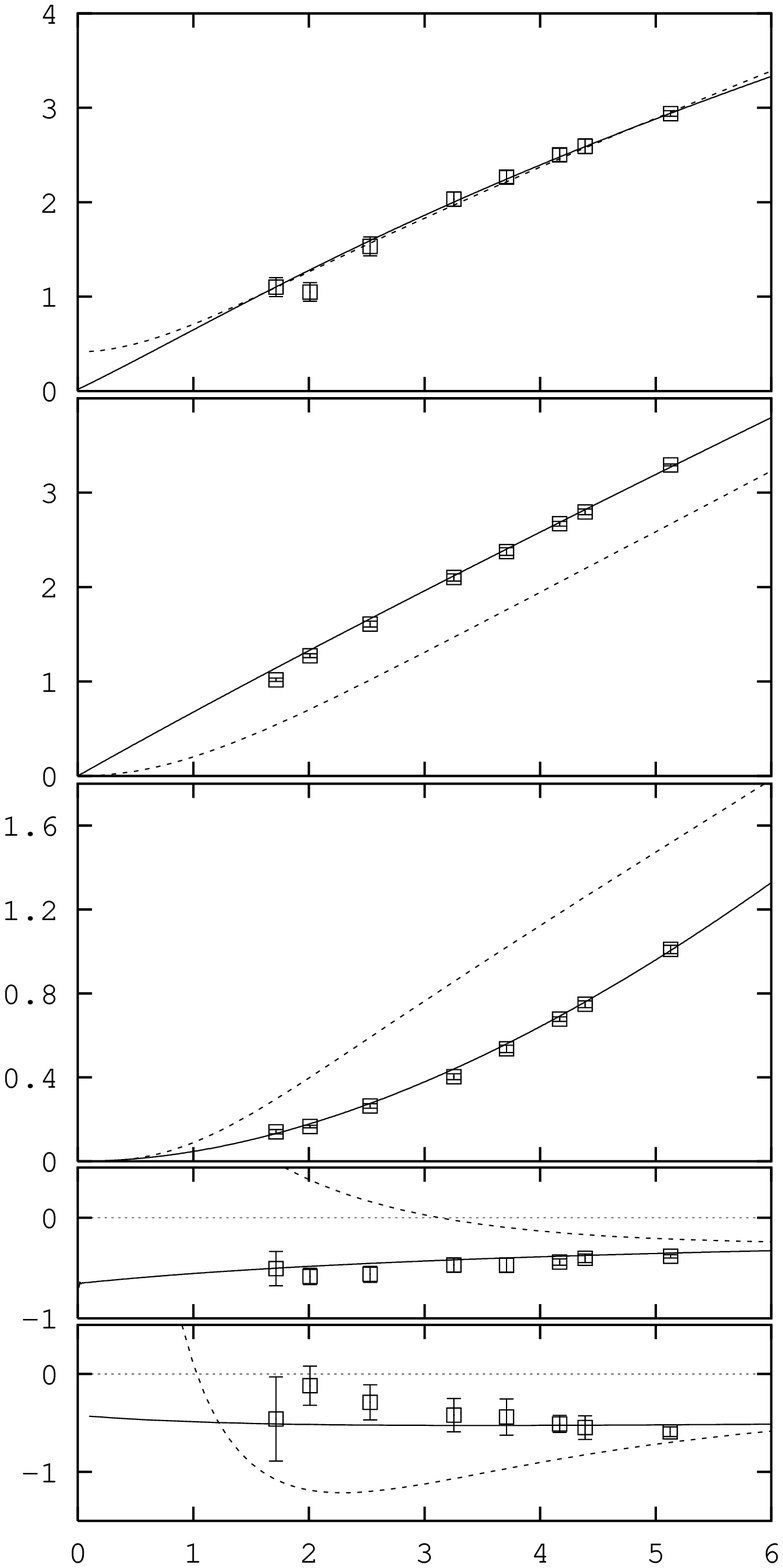,bbllx=4.5cm,bblly=0.1cm,bburx=18.cm,bbury=27.cm,height=16cm}}
}          \end{center}
\mbox{\hspace{6.4cm} $Y = \log (\sqrt{s}/2 Q_0)$}
\vspace{0.3cm}
\fcaption{The average multiplicity $\bar {\mathcal N}_E$ and the first four
cumulants of charged particles' energy spectra $E dn/dp$ vs. $\xi_E$,i.e.,
the average
value $\bar \xi_E$, the dispersion $\sigma^2$, the skewness $s$ and the
kurtosis $k$,
are shown as a function of $Y = \log (\sqrt{s}/2Q_0)$ for $Q_0$ = 270 MeV.
Data points from MARK I\protect\cite{slac} at $\sqrt{s}$ = 3, 4.03, 7.4
GeV, TASSO\protect\cite{tasso}  at $\sqrt{s}$ = 14, 22, 35, 44
GeV and LEP at $\sqrt{s}$ = 91.2 GeV (weighted averages of
ALEPH\protect\cite{aleph}, DELPHI\protect\cite{delphi}, L3\protect\cite{l3} and
OPAL\protect\cite{opal}).
Predictions of the Limiting Spectrum (i.e. $Q_0 = \Lambda$)
of MLLA with running $\alpha_s$
 and of MLLA with fixed $\alpha_s$ are also shown (for $n_f$ = 3).
Predictions of the average multiplicity refer to the two-parameter formula
$\bar {\mathcal N}_E = c_1 2 \frac{4}{9} \bar {\mathcal N}_{Part} + c_2$.}
\label{fig1}
\end{figure}
For the average multiplicity we added two new free parameters:
\be
\bar \N_E = c_1 \frac{4}{9} 2 \bar \N_{LS} + c_2
\ee
where
the factor $4/9$ is the proportionality factor between quark and gluon jets
and  the factor 2 accounts for the two hemispheres; the two additional free
parameters $c_1$ and $c_2$ give the proportionality factor between
parton and hadron spectra  and  the leading particle contribution
respectively. This second parameter is needed to recover the right behavior
near threshold, as the prediction of the Limiting Spectrum does not fulfill
the correct boundary conditions.  The values of the two parameters
have been fixed by fitting the lowest and the highest energy data points.

The predictions from MLLA with running $\alpha_s$ and normalization at
threshold are remarkably successful
considering the fact that there are only two parameters, actually coinciding,
 for the four cumulants. The deviation in $\bar \xi_E$ at the lower energies
 may suggest that the leading valence quark, which is  now
 neglected by restricting to the $D^+$ spectrum, plays indeed a role at low
c.m.
 energies.  Otherwise, the predictions are confirmed at an almost
quantitative level down to the lowest c.m. energies.

Figure~1  also shows the
theoretical predictions for MLLA with fixed $\alpha_s$
(dashed lines) for the same value of $Q_0$ and $n_f$ and
$\gamma_0$ = 0.64.
This value of  $\gamma_0$ has been chosen in order to obtain
a good fit for the average multiplicity, where we proceed as above
to a fit with two additional free parameters.
With this  value of $\gamma_0$,
the asymptotic slope for $\bar \xi_E$ is also  well reproduced. An adjustement
of the absolute normalization of  $\bar \xi_E$
at a particular energy $Y_0$ would be
possible if the perturbative QCD
evolution towards lower energies and the normalization at
threshold are abandoned.
Looking at higher order moments, however, the model with fixed $\alpha_s$
is unable to reproduce the experimental behavior. In addition,
the fixed $\alpha_s$ regime can be excluded already close to threshold.
Indeed, if we suppose that the coupling is
fixed for a certain energy interval near threshold $Y_0 \le Y \le Y_1$ and only
runs for $Y > Y_1$, then $\bar \xi$ would be shifted towards smaller values
because of the very different evolution near threshold,
in contrast to the experimental behavior.

\subsection{Rescaled cumulants}

In  addition to the standard moment analysis, it is interesting to
consider the rescaled cumulants  $\kappa_q/\bar \xi$;
in the high energy limit
they become energy independent in the model with fixed coupling (see
eq.~\eref{mom2}) and therefore
exhibit more directly the difference to the case of running coupling
in the asymptotic domain.

\begin{figure}
\begin{center}
\mbox{\epsfig{file=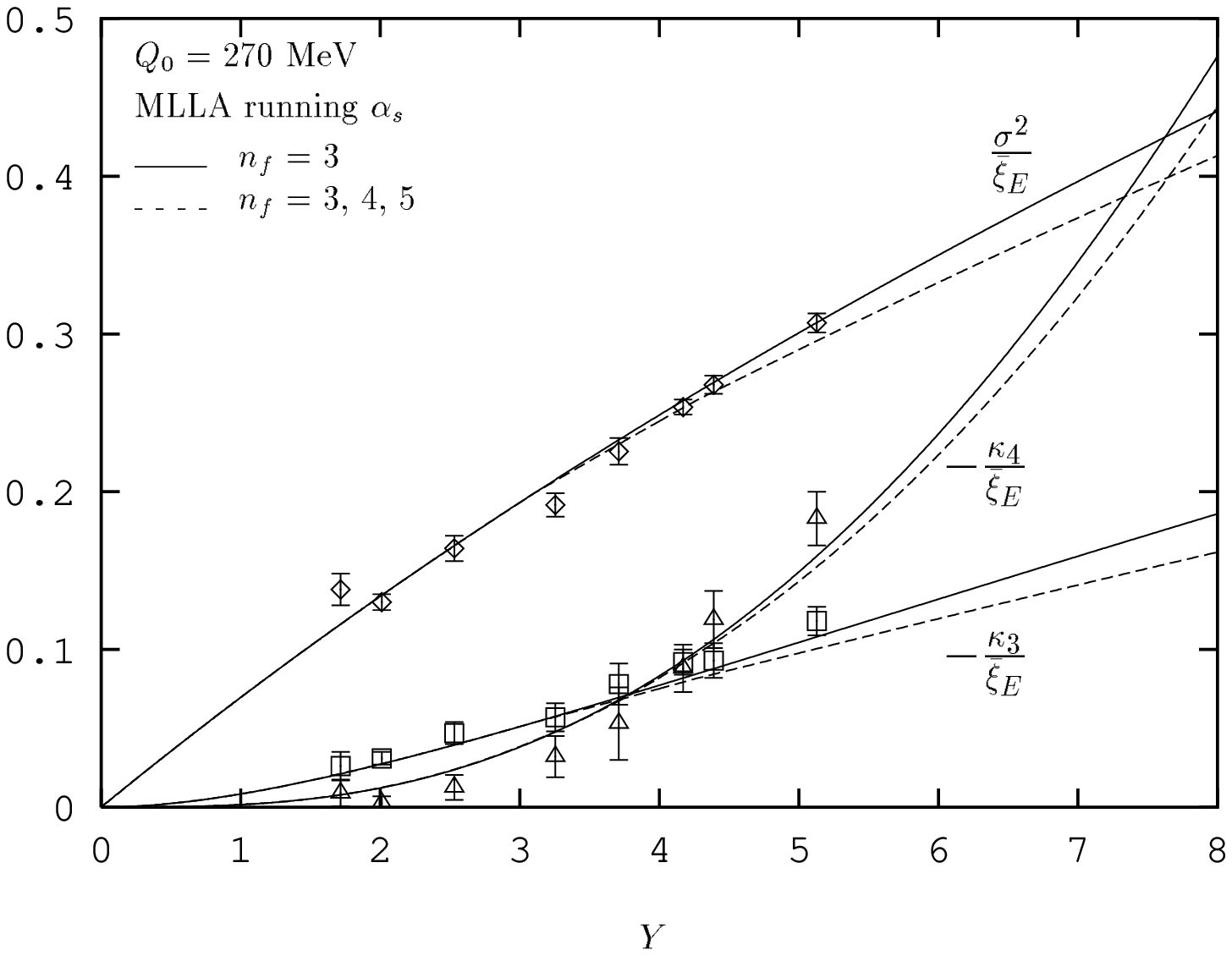,height=7cm,width=11cm,bbllx=4.cm,bblly=8.cm,bburx=19.cm,bbury=18.cm}}
\end{center}
\fcaption{Rescaled cumulants $\kappa_q/\bar \xi$
as a function of $Y = \log (\sqrt{s}/2 Q_0)$ with the corresponding
predictions of the Limiting Spectrum of MLLA
with running $\alpha_s$ either without or with heavy flavors included.
Data as in Figure~1.
For fixed $\alpha_s$ these quantities approach constant values at high
energies.}
\label{fig2}
\end{figure}

Figure~2  show the experimental results on rescaled cumulants of order 2, 3 and
4 for $Q_0$ = 270 MeV;
a clear  $Y$-dependence in the available energy range is visible;
this result is well reproduced by predictions of MLLA with running $\alpha_s$
and $n_f$ = 3, while is again in contradiction with the constant behavior
expected for fixed $\alpha_s$.
It would be interesting to continue this type
of studies for instance at TEVATRON,  where jets of
higher energies are available, in order to test whether the rescaled cumulants
show the remarkable dependence on $Y$  predicted by the theory.

Let us also comment the dependence of the above results on the number of active
flavors $n_f$.
In leading order $n_f$ enters only through the running coupling
$\alpha_s(Y,n_f)$,  while at next-to-leading level
also through the parameter $a$. It has been argued that the
number of flavors $n_f$ should be kept fixed at  3,
as light quarks should dominate quark-pair
production in a well-developed cascade.
However, one can also include heavy quark thresholds in the calculation
simply by taking $\gamma_{\omega}(y)$ under the
integral with $n_f$ as the number of open flavors at energy $y$.
Restricting our discussion to the leading term we note that in the original
eq.~\eref{uno} the scale of $\alpha_s$ is $k_t \simeq z(1-z) P < P/4$.
Therefore,  we included the effect of heavy quarks
in $n_f$ for momentum $P > 4 m_Q$.
The effect  of including heavy quarks thresholds in the model gives a
correction of a few \%, as shown for rescaled cumulants
in Figure~2 (dashed lines). In addition, let us remind that
further uncertainties
on the scale of the running, two-loop effects  and a possible
explicit mass dependence of the coupling\cite{DKTnew} are neglected in this
approach.

\subsection{Low-energy behavior: Boltzmann factor or running-$\alpha_s$ effect}

Let us consider  again the
Lorentz-invariant distribution $E dn/d^3p$  as a function of the particle
energy $E$ for  small energies up to 1 GeV.
The experimental exponential behavior is often related to the
Boltzmann factor in the thermodynamical
description of multiparticle production.

\begin{figure}
\vfill \begin{minipage}{.48\linewidth}
          \begin{center}
   \mbox{
\mbox{\epsfig{file=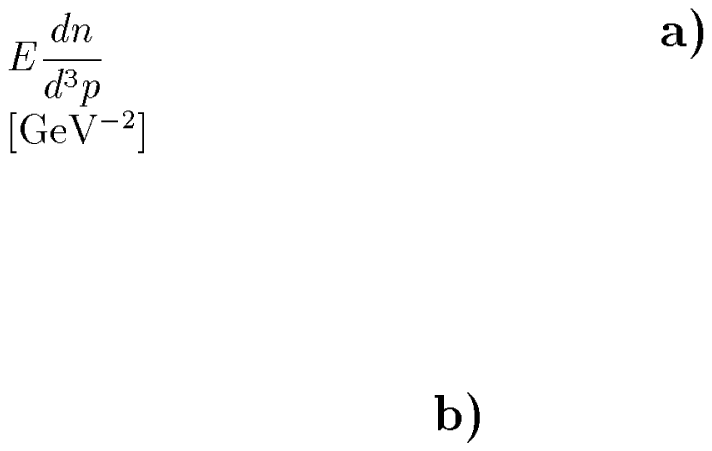,bbllx=4.4cm,bblly=17.5cm,bburx=4.6cm,bbury=24.cm}}
\mbox{\epsfig{file=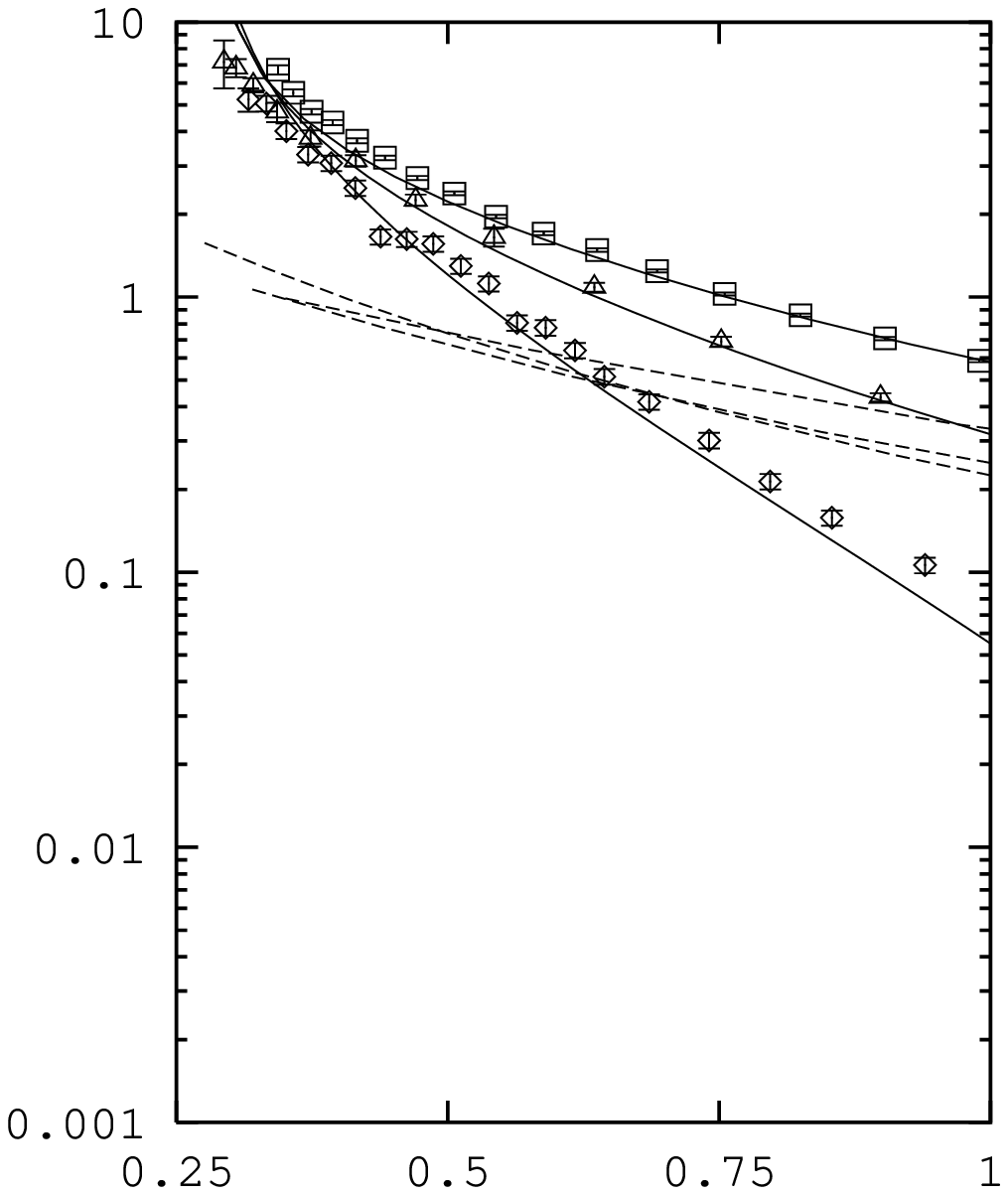,width=.72\linewidth,bbllx=4.5cm,bblly=2.5cm,bburx=12.cm,bbury=13.cm}}
\hspace{-5.5cm}\mbox{\epsfig{file=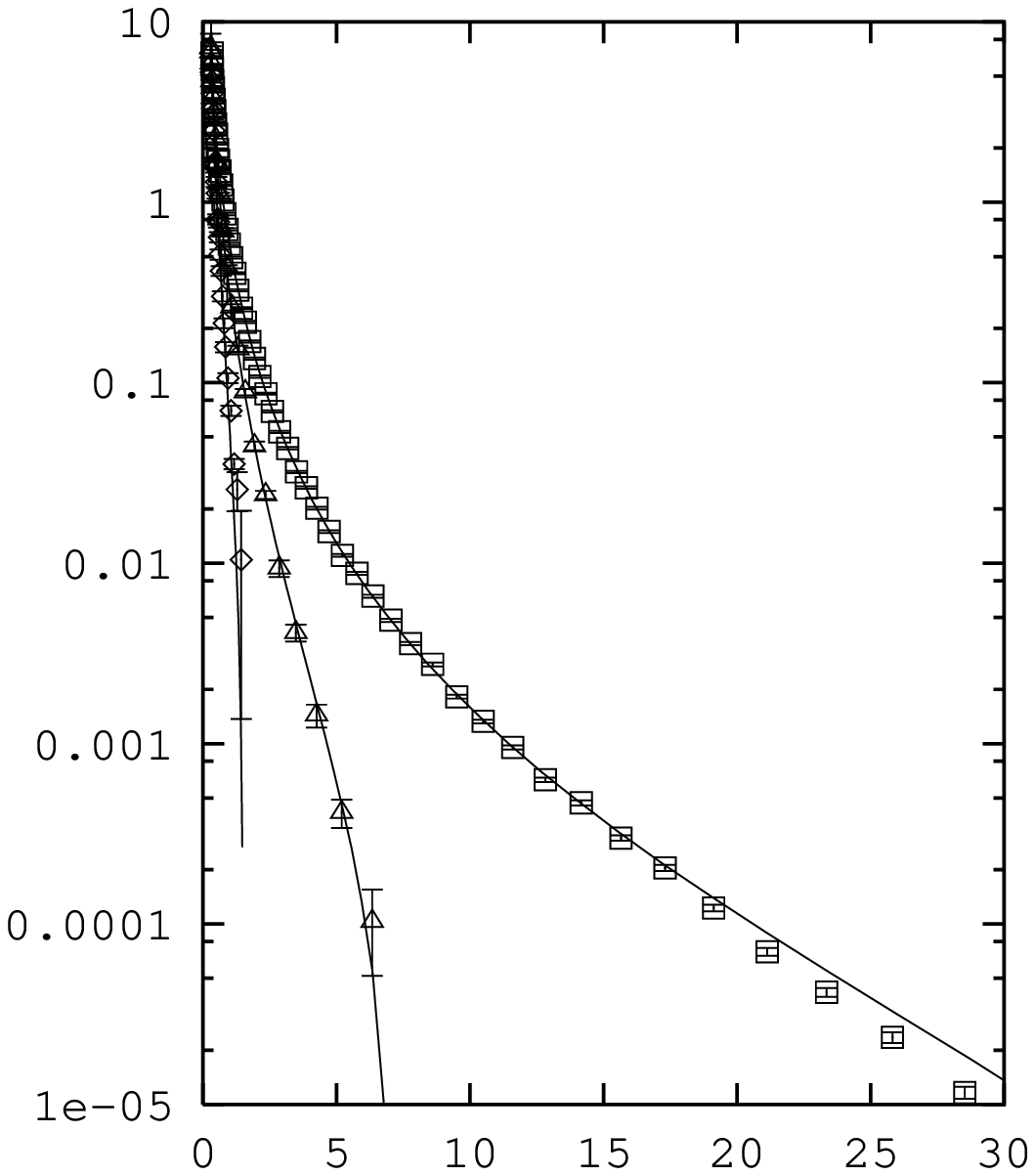,width=.7\linewidth,bbllx=1.cm,bblly=0cm,bburx=16.cm,bbury=15.cm}}
  }        \end{center}
\centerline{$\quad \qquad E$ [GeV]}
      \end{minipage}\hfill
      \begin{minipage}{.48\linewidth}
          \begin{center}
   \mbox{
\mbox{\epsfig{file=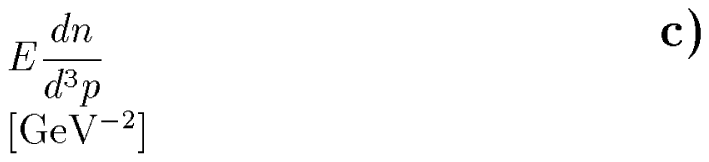,bbllx=4.4cm,bblly=17.5cm,bburx=4.6cm,bbury=24.cm}}
\mbox{\epsfig{file=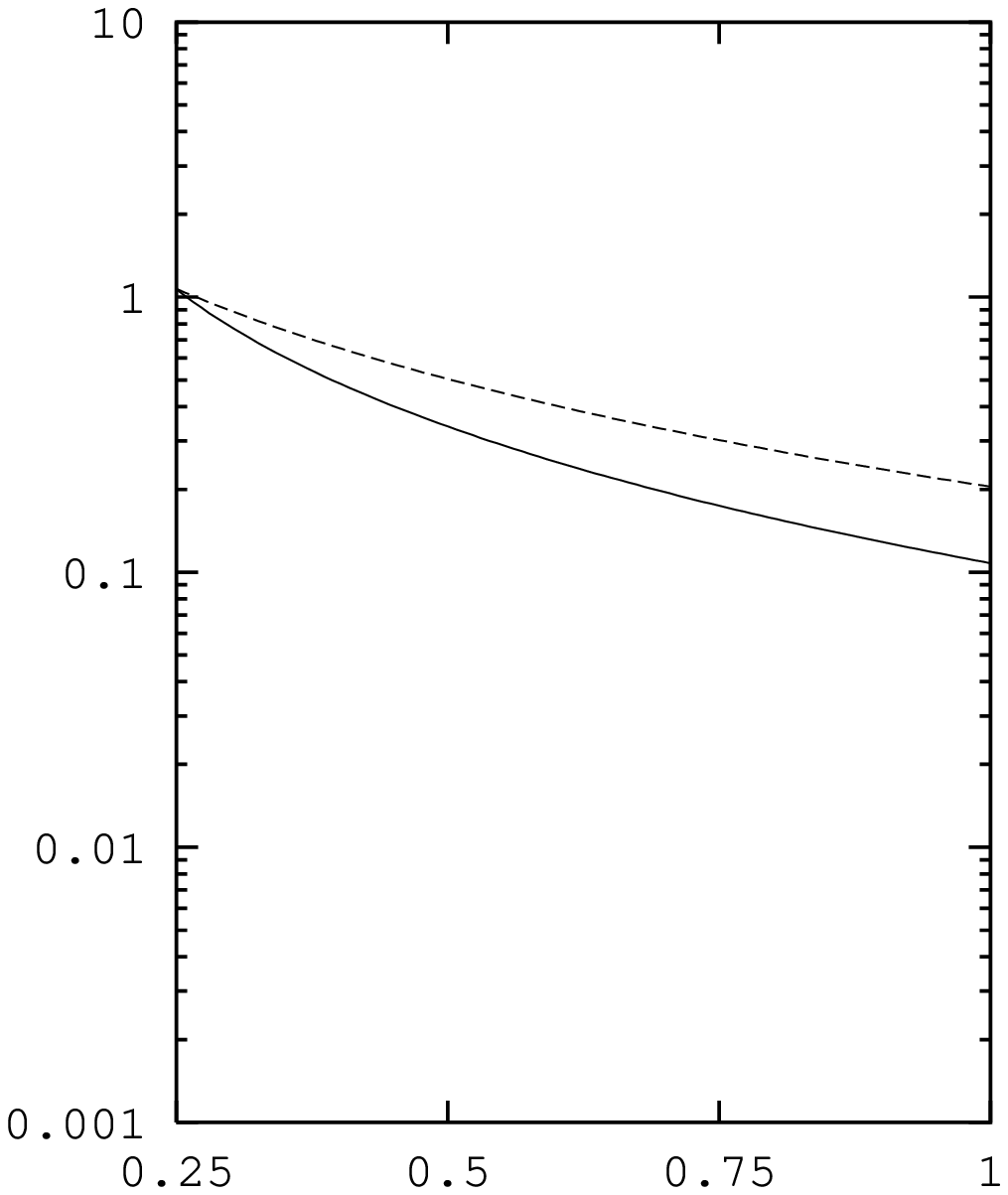,width=.72\linewidth,bbllx=4.5cm,bblly=2.5cm,bburx=12.cm,bbury=13.cm}}
  }        \end{center}
  \centerline{$\quad \qquad E$ [GeV]}
      \end{minipage}
\fcaption{Invariant distribution $E dn/d^3p$
as a function of the particle energy $E$ for $Q_0$ = 270 MeV.
{\bf a)} comparison of experimental data from MARK I\protect\cite{slac}
(diamonds),
TASSO\protect\cite{tasso} (triangles) and OPAL\protect\cite{opal}  (squares)
with  predictions from MLLA with running $\alpha_s$ (Limiting Spectrum)
(solid lines) and with fixed $\alpha_s$ (dashed lines).
{\bf b)} same  as in {\bf a)}, but  in the energy range up to 30
GeV; only predictions from
MLLA with running $\alpha_s$ are shown; {\bf c)} analytical predictions of the
Born term of DLA with running
$\alpha_s$ (solid line) and with fixed $\alpha_s$ (dashed line); both curves
are arbitrarily normalized to a common value.}
\label{fig3}
\end{figure}

Figure~3a shows experimental data  on the invariant distribution at three
different c.m. energies. All three
distributions  tend to a common behavior at very low energies $E$, even though
deviations from a simple exponential are visible at larger c.m. energies.
Theoretical predictions for the Limiting Spectrum of MLLA with
running $\alpha_s$ and for MLLA with fixed $\alpha_s$ (with $\gamma_0$ = 0.64)
with the same value of $Q_0$ are also shown.
Both theoretical curves are normalized to the experimental average
multiplicity.
Figure 3b compares the same data with theoretical predictions from the
Limiting Spectrum  of MLLA with running $\alpha_s$ in a wide energy range
up to 30 GeV.
It is remarkable that the Limiting Spectrum of MLLA  is
in good agreement with experimental
data in a wide energy range and can approximately reproduced
 the exponential decrease of the spectrum in the low energy region.
On the contrary, the exact solution of
the fixed-$\alpha_s$ model shows a flatter behavior,
inconsistent with experimental behavior.

This behavior near the kinematical limit $E \to Q_0$ can be qualitatively
understood in an analytical way in the DLA, where eq.~\eref{uno} becomes:
\be
D(\xi,y) = \delta(\xi) + \int_0^{\xi} d\xi' \int_0^y dy' \frac{C_a}{N_c}
\gamma_0^2(y') D(\xi',y')
\ee
where $y = \log k \Theta / Q_0 = \log k_T/Q_0 = Y-\xi$.

The above  evolution equation can be solved iteratively;  with one iteration,
one gets the Born term, which is the leading term in the limit $E \to Q_0$:
\be
D(\xi,y) = \delta(\xi) + \frac{C_a}{N_c} \int_0^y dy' \gamma_0^2(y')
\ee
The difference between the fixed and the running $\alpha_s$ case is fully due
to the expression of the anomalous dimension:
in the fixed-$\alpha_s$ model, $\gamma_0$ becomes a constant, then:
\be
D_{fix}(\xi,Y) = \delta(\xi) + \frac{C_a}{N_c} \gamma_0^2 (Y-\xi)
\ee
whereas for running $\alpha_s$, $\gamma_0^2(y') = 12/b/(y'+\lambda)$,
\be
D_{run}(\xi,Y) = \delta(\xi) + \frac{C_a}{N_c} \frac{12}{b} \log \left(
\frac{Y-\xi+\lambda}{\lambda} \right)
\ee
Notice that in both cases the spectra go to finite values independent of
c.m. energy in the limit $E \to Q_0$; the dependence of the spectrum on c.m.
energy enters in higher orders terms.
Predictions for the Born terms of DLA in the two models (with arbitrary
normalization) are shown in Figure~3c; notice
the relative enhancement at low energy $E$ in the model with running
$\alpha_s$. This behavior can be intuitively explained:
as for decreasing particle energy $E$ also the typical
particle $p_t$ is
necessarily decreasing, the coupling $\alpha_s(p_t/\Lambda)$ is rising in the
running $\alpha_s$ case, whereas the fixed $\alpha_s$ model cannot account for
such an effect.

\section{Conclusions}

The analysis of energy spectra in terms of moments has been performed. A
consistent kinematical scheme to  relate parton and hadron spectra in
the low-energy region has been defined. This yields
a good agreement of experimental data with theoretical predictions of MLLA
with running $\alpha_s$ plus LPHD in a wide c.m. energy range from 3
GeV  up to LEP energy.
The best agreement with
data is found for the Limiting Spectrum, where the two parameters of the theory
coincide, i.e., $Q_0 = \Lambda$; the best
value of $Q_0$ has been estimated to be 270$\pm$20 MeV.
The zero-th order moment, i.e., the average multiplicity, requires two more
parameters to fix the overall normalization for particle production and the
leading particle contribution.
Predictions of MLLA plus LPHD but with fixed coupling
are inconsistent with the experimental behavior in the full energy range, thus
showing a direct evidence for the running of the coupling in hadronic energy
spectra.
The effect of the running $\alpha_s$
is most pronounced near threshold where the variation of the coupling is
strongest; it approximately reproduces the exponential dependence of
the Lorentz-invariant distribution $E dn/d^3p$
at low particle energies $E$ of few hundreds MeV.
The analysis of rescaled cumulants is shown to be more sensitive to the
difference with the fixed-$\alpha_s$ regime in the high-energy region, for
instance at TEVATRON energies.
A similar analysis ranging from low to high values of the available energy
 can be performed
also at HERA by looking at hadronic jet production in the Breit frame.

The obtained results  give further support to the LPHD picture.
A word of caution is appropriate nevertheless. The MLLA is
based on high energy approximations
(restriction to $D^+$ contribution, neglect of
Next-to-Next-to-Leading effects) which have been extended into the low-energy
region. A deviation in $\bar \xi$ at
3 GeV could  indeed suggest that further contributions are needed.
Also additional scheme-dependence of the coupling and two loop effects
have been neglected in the calculation. Finally let us remind
that we safely identified $Q_0$ with an effective hadron mass, since we
referred
to single inclusive spectra for all charged particles, but
this identification  cannot be pushed forward to single inclusive spectra
for identified
particles without taking into account the effects of resonance decay,
as recently shown in \cite{deangelis}.

\section{Acknowledgements}

I am grateful to  Wolfgang Ochs for fruitful suggestions on the content of
this note.
I would like to thank Ladislav \v Sandor  and the Organizing Committee
for the nice  atmosphere created at this Symposium.
Financial support from D.A.A.D. is gratefully acknowledged.

\end{document}